\documentclass{article}

\usepackage{PRIMEarxiv}

\usepackage[utf8]{inputenc}
\usepackage[T1]{fontenc}
\usepackage{hyperref}
\usepackage{url}
\usepackage{booktabs}
\usepackage{amsfonts}
\usepackage{nicefrac}
\usepackage{microtype}
\usepackage{lipsum}
\usepackage{fancyhdr}
\usepackage{graphicx}
\usepackage{multirow}
\graphicspath{{media/}}
\usepackage{threeparttable}

\title{AVI-Personality: A Trait-Activated Multimodal Dataset for Personality and Competency Assessment in Asynchronous Video Interviews}

\author{
Tianyi~Zhang \\
Southeast University \\
\texttt{t.zhang@seu.edu.cn}
\And
Jinwenxi~Shang \\
Southeast University \\
\And
Antonis~Koutsoumpis \\
Vrije Universiteit Amsterdam \\
\AND
Yuan~Zong$^{*}$ \\
Southeast University \\
\And
Reinout~E.~de~Vries \\
Vrije Universiteit Amsterdam \\
\And
Wenming~Zheng \\
Southeast University \\
}

\begin{document}
\maketitle

\begin{abstract}
With the rapid development of AI-based personality and job-related competency assessment, Asynchronous Video Interviews (AVIs) are increasingly used in recruitment. However, existing multimodal personality datasets are often based on short, task-free social media videos and crowdsourced apparent personality labels, which limits their construct validity and relevance to structured interview assessment. To address these limitations, we introduce AVI-Personality, a trait-activated multimodal dataset for personality and job-related competency assessment from AVIs. The dataset contains 3,876 interview videos from 646 participants who completed a simulated management traineeship application. Participants answered two generic questions and four personality-targeted questions designed according to Trait Activation Theory. Our dataset provides both self and observer-reported HEXACO personality traits and job-related competency. We validate AVI-Personality through reliability, construct validity, internal nomological association, fairness, and benchmark analyses. Validation results show that the observer-rated personality traits have moderate to high reliability, especially when ratings are based on personality-targeted questions. Benchmark results show that text-based AI algorithms provide strong personality-relevant cues, while multimodal methods achieve the best overall performance but only modestly outperform text-based baselines. In general, AVI-Personality provides a psychometrically grounded dataset for developing and evaluating AI-based models for personality and competency assessment. The dataset is available are released at \href{https://github.com/APAL-SEU/AVI6}{\textit{https://github.com/APAL-SEU/AVI6}}.
\end{abstract}

% keywords can be removed
\keywords{multimodal dataset, personality recognition, trait activation, asynchronous video interviews }

\section{Introduction}
%\IEEEPARstart{E}{mployment} interviews remain a fundamental tool for evaluating candidates across a wide range of industries \cite{Huffcutt2011, Levashina2014}.
%Traditional face-to-face interviews rely on subjective judgment and have limited scalability. Through standardized questioning and scoring processes, structured interviews are superior to unstructured interviews in predicting job performance \cite{Campion_Palmer_Campion_1997, Huffcutt2011}. 

Asynchronous Video Interviews (AVIs) allow interviewees to
record responses to predefined questions remotely. In recent years, they have become a common method in remote recruitment due to their evaluation consistency and large-scale screening capabilities \cite{ Suen2019}. Automated Personality Trait Recognition (APTR) is a key technical approach in the AVI scenario. This technology assesses personality traits based on candidates’ verbal content, vocal features, and nonverbal behavioral signals in videos. Existing studies have shown that personality traits play an important role in predicting job performance and person–job fit \cite{Barrick_Mount_1991, McCrae2003}. Thus, the growing demand for scalable and efficient candidate screening has increased interest in AI algorithms for automatically assessing personality traits and job-related competencies.

%makes the automatic modeling of multi-source behavioral signals in interviews possible, and promotes the transformation of traditional face-to-face interviews toward a data-driven AVI paradigm \cite{Suen2019}. This paradigm shift makes the evaluation process more scalable and cost-efficient, and provides opportunities to apply machine learning techniques for multimodal analysis \cite{Poria2017}. 

These advances, however, rely on high-quality AVI datasets that provide multimodal behavioral signals and reliable annotations which enables robust model development and evaluation. Thus, several multimodal datasets have been collected for personality and job-related competencies assessment. For example, the ChaLearn First Impressions dataset \cite{PonceLopez2016, Junior2019} contains short video clips of individuals from online videos and provides annotations of apparent personality traits based on the Big-five personality Model. Its extension dataset, ChaLearn First Impressions V2 \cite{Escalante2020Modeling}, extends the annotation scope by including an additional interview-performance score, which reflects whether a candidate is likely to be invited for a job interview.

%APTR has made significant progress in the AVI domain, in which multimodal datasets have played a central supporting role.
%Several benchmark corpora have been proposed to facilitate the development of machine learning models for personality assessment based on audiovisual data.
%Among these datasets, the ChaLearn First Impressions dataset \cite{PonceLopez2016, Escalera2018} has become a widely adopted benchmark. 
%This dataset contains short video clips of individuals from online videos and provides annotations of apparent personality traits based on the Five-Factor Model, and has been widely used for the development of models and algorithms that learn personality representations from visual and vocal behaviors. These datasets have driven significant progress in multimodal representation learning and predictive modeling, and have shown that personality assessment can be inferred from short-term behavioral observations.

Although previous works have collected several multimodal datasets for personality traits and job-related competencies assessment, most of these datasets rely on publicly available video data. Such datasets are often sourced from open-domain video content unrelated to interview tasks (e.g., YouTube), with highly diverse topics and a lack of task constraints.
According to Trait Activation Theory (TAT) \cite{tett2003personality}, accurate measurement of personality traits depends on situational tasks that can activate the relevant traits.
Task-irrelevant data sources weaken the validity of personality traits and interview performance assessment.

Beyond limitations in the input videos, the observer annotations in these datasets are typically characterized using single bipolar adjectives (e.g., Friendly vs. Reserved for Extraversion in the ChaLearn First Impressions dataset),  or directly provided as overall perceived personality scores. In psychological research, personality constructs are usually measured using multiple items or Behavioral Anchored Rating Scales (BARS) \cite{Ashton2007}. Such simplified annotation schemes cannot cover the full construct of personality and reduce construct validity of annotation. In addition, annotators in these datasets are typically recruited from crowdsourcing platforms (e.g., Amazon Mechanical Turk) and lack professional training in personality assessment tasks. Previous studies have shown that training can significantly improve measurement consistency and accuracy \cite{Buhrmester_Kwang_Gosling_2016}. Thus, the lack of training among annotators undermines the effectiveness of automatic interview assessment.

To address these limitations, this paper introduces AVI-Personality, a multimodal AVI dataset for personality and job-related competencies assessment. Our dataset contains 3,876 videos from 646 participants who completed mock interview in a simulated job application procedure, which provides a standardized and valid interview context rather than short in-the-wild social media videos. The questions in AVIs consist of two generic selection questions and four trait-related personality questions that target HEXACO traits. Thus, the input videos support trait activation and real-world recruitment analysis. The dataset also provides psychologically grounded annotations. The observer-reported personality traits are annotated through BARS-based ratings by trained psychologists. The job-related competencies cover recruiter-rated interview performance and cognitive ability. These designs improve both the validity of input videos and the construct validity of annotations. Therefore, our dataset can support robust model development for AVI-based personality and competency assessment. In general, our work makes following contributions: 

\begin{itemize}%[leftmargin=0.3cm]
\item Different from prior datasets based on short in-the-wild social media videos, AVI-Personality provides structured interview responses elicited by generic selection questions and trait-related personality questions. Our dataset also includes self-reported and observer-reported personality traits and job-related competencies. This combination moves AVI assessment beyond superficial first impressions by grounding behavioral evidence in theoretically relevant situations and psychologically grounded annotations.

\item We validate the annotation of our dataset from the perspectives of reliability, construct validity, internal nomological association, and demographic fairness. The analyses show that the annotations are not only consistent across trained raters, but also meaningfully connected to self-reported traits and recruitment-relevant outcomes. These results provide psychometric evidence for the quality and practical relevance of the dataset. 

\item  We conduct a comprehensive benchmark on our dataset using text-based, audio-based, visual-based, and multimodal-based AI modals. The results show that text-based methods provide strong personality-relevant cues, while multimodal methods achieve the best overall performance but only modestly outperform strong text-based baselines. The benchmark establishes reference results for AVI-based personality assessment and reveals the relative contribution of different behavioral modalities.
\end{itemize}

%To improve construct validity and measurement richness, the proposed dataset integrates video, audio, and textual modalities.It provides multiple complementary annotations, including self-reported personality, observer-rated personality, observer-rated job performance and intelligence. %认知水平的说法
%In contrast to prior work, the interview protocol is explicitly designed based on Trait Activation Theory to elicit diagnostically relevant behaviors for the HEXACO personality dimensions. 
%In addition, data collection is conducted in a controlled, simulated interview setting, ensuring stronger ecological validity compared to datasets derived from unconstrained online videos. 
%Finally, all annotations are provided by trained raters and experienced recruitment professionals, resulting in improved reliability and closer alignment with real-world evaluation practices.

\section{Related Work}

In this section, we first discuss personality traits and job-related competencies as the psychological constructs to be assessed. We then review automatic assessment methods that predict these constructs from behavioral signals in video interviews. At last, we summarize existing datasets that support these tasks and discuss their limitations.

\subsection{Personality Traits and Job-related Competencies}

The theoretical foundation of automated personality trait recognition (APTR) relies on robust psychometric models and reliable behavioral measurements \cite{Zhao2022, Vinciarelli2014, AIReview2025}. Among existing personality frameworks, the Five-Factor Model (Big Five) and the HEXACO model are the two most widely adopted approaches in personality and organizational behavior research. The Big Five personality model describes personality through five dimensions: extraversion, conscientiousness, agreeableness, neuroticism, and openness to experience \cite{articleMcCrae2003}. This model has been extensively validated in cross-cultural studies and various workplace-related studies \cite{Levashina2014}. Compared with the Big Five personality model, the HEXACO model additionally introduces the Honesty-Humility dimension to characterize sincerity, fairness, modesty, and greed avoidance \cite{Ashton2007}. Existing organizational behavior studies have shown that HEXACO provides incremental validity in predicting workplace deviance, integrity-related behaviors, and organizational citizenship behavior \cite{Anglim_Lievens_Everton_Grant_Marty_2018, Oh_Le_Whitman_Kim_Yoo_Hwang_Kim_2014}.

Job-related competencies refer to observable work-relevant capabilities that reflect how well a candidate can meet the behavioral requirements of a target position \cite{heijke2003fitting}. In personnel selection, competencies are often used to complement personality traits since they describe more specific and job-oriented behavioral outcomes\cite{bartram2005great,campion2011doing}. For AVI-based assessment, several competencies are particularly important, including integrity, collegiality, social versatility, and development orientation. For example, integrity reflects ethical conduct and trustworthiness in workplace interactions and collegiality captures cooperation and willingness to support others \cite{Koutsoumpis2024}. 

In interview and personnel selection scenarios, HEXACO personality model demonstrates particularly strong explanatory value. For example, conscientiousness is closely related to task organization ability, responsibility, and job performance \cite{Koutsoumpis2024,bartram2005great}. In addition, extraversion is associated with communication effectiveness and social interaction while honesty-humility is highly related to ethical behavior and integrity in organizations \cite{Judge2002, Oh_Le_Whitman_Kim_Yoo_Hwang_Kim_2014}. These personality traits are usually behaviorally expressed during interview interactions, making them particularly important for AVI-based personality and competency assessment.

%Although APTR research has achieved substantial progress, existing studies still have several limitations in workplace-oriented personality assessment. Many datasets are collected from unconstrained social media videos or “in-the-wild” recordings, where behavioral expressions are natural but weakly related to workplace-relevant competencies \cite{PonceLopez2016, Hickman2022}. Therefore, these behaviors cannot stably reflect workplace-related personality manifestations.

%According to Trait Activation Theory (TAT), personality traits are expressed as behavioral responses to situational cues \cite{Tett2003}. Different interview questions or task settings selectively activate different personality-related behaviors \cite{Levashina2014}. For example, self-introduction questions may more easily elicit behaviors related to extraversion, while ethical dilemma questions are more likely to activate honesty-humility and emotion regulation related behaviors. Therefore, structured asynchronous video interviews (AVIs), which combine standardized questions with behavior-eliciting situations, provide a more suitable environment for observing workplace-related personality manifestations.
\subsection{Automatic Personality and Competency Assessment}
Personality and competency assessment plays an important role in personnel selection, organizational psychology, and human resource management \cite{Huffcutt2011, Levashina2014}. Traditional assessment methods usually use self-reported questionnaires. Although questionnaire-based evaluation has strong interpretability and psychometric validity, it is vulnerable to social desirability bias, as candidates may intentionally present themselves in a more favorable way during personnel selection. This response distortion can reduce the authenticity of self-reported personality scores and weaken their practical value in high-stakes recruitment contexts. To address these limitations, AI-based automatic personality and competency assessment methods have attracted increasing research attention in recent years.

Traditional automatic personality and competency assessment methods are mainly based on machine learning models and hand-crafted features. Early studies usually extracted low-level behavioral features from video, speech, and text, such as facial expressions, head movements, speech rate, pitch, and lexical statistical features, and combined them with traditional machine learning methods such as Support Vector Machines (SVM), Random Forests, or linear regression for personality prediction \cite{Vinciarelli2014, Mehta_Majumder_Gelbukh_Cambria_2020}. With the development of deep learning technics, research focus gradually shifted from hand-crafted features to end-to-end representation learning methods. Convolutional Neural Networks (CNNs), Recurrent Neural Networks (RNNs), and Transformer-based models have been widely used to automatically learn personality-related representations from multimodal data \cite{Suen_Hung_Lin_2019, Giritlioglu_Mandira_Yilmaz_Ertenli_Akgur_2021, Zhao2022}.
In recent years, Large Language Models (LLMs) and Multimodal Large Language Models (MLLMs) have started to be applied to personality assessment tasks. Related study \cite{wright2026generative} shows that LLMs can extract high-level semantic features from open-ended language and interview transcripts to improve personality assessment through psychology-informed prompting %\cite{Li_He_Xu_Luo_Hu_Hong_Wang_2025, Zhu2025}.

The performance of these algorithms strongly depend on the availability of high-quality multimodal datasets. Previous works \cite{Vinciarelli2014, Junior2019, Zhao2022} have shown that the reliability of personality labels and the richness of behavioral modalities can affect model validity, interpretability, and generalizability for personality assessment. Similarly, in automatic interview and job-related competency assessment, structured interview design and reliable human ratings are essential for ensuring that computational scores reflect meaningful work-related constructs rather than superficial behavioral cues \cite{Hickman2022}. Thus, dataset construction has become a core issue in APTR and competency assessment research.

\begin{figure*}
    \centering
    \includegraphics[width=1\linewidth]{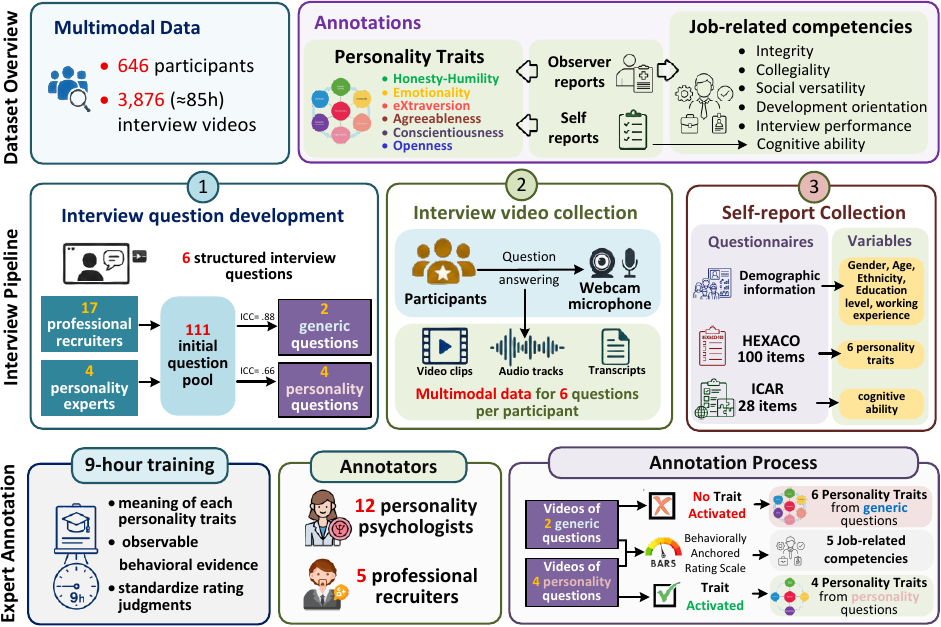}
    \caption{Overview of the AVI-Personality dataset. Guided by Trait Activation Theory, the dataset uses structured asynchronous interview questions to elicit personality-relevant behaviors. Expert annotations were provided by trained professional recruiters and personality psychologists using standardized rating criteria.}
    \label{fig:data_set}
    
\end{figure*}

\subsection{Datasets for Personality and Competency Assessment}
The rapid development of APTR research has promoted the emergence of numerous multimodal personality and interview datasets. For example, the ChaLearn First Impressions dataset provides short video clips annotated with apparent Big Five personality traits \cite{PonceLopez2016}. The UDIVA dataset contains dyadic interaction videos with multimodal behavioral signals and personality annotations \cite{Palmero2021}. For interview-based dataset, Hickman et al. \cite{Hickman2022} collected a video-based dataset which supports the validation of AVI-based personality assessment in a mock interview setting.

Many widely used datasets, such as the ChaLearn First Impressions dataset \cite{PonceLopez2016}, are based on short social media video clips and rely on crowd-sourced apparent personality annotations. These datasets provide large-scale multimodal resources and facilitate the development of data-driven APTR models. However, their behavioral scenarios usually lack structured design and their annotations mainly reflect first impressions rather than validated personality traits \cite{Oh2011}. Interview-oriented datasets, such as MIT Interview \cite{Hoque2013}, introduce more structured behavioral environments and task-driven interactions. However, most existing interview datasets still have several limitations. Their annotation procedures largely rely on crowd ratings without professionally validated psychometric instruments. Thus, the annotated labels capture general impressions rather than theoretically grounded personality traits or job-related competencies. In addition, most datasets lack joint annotations of self-reports and multi-rater observer ratings. This limitation makes it difficult to examine self–other agreement and validate whether automatic models reflect stable psychological constructs rather than surface-level behavioral cues.

Compared with previous datasets, our dataset introduces AVI questions designed based on Trait Activation Theory \cite{tett2003personality} to elicit specific behavioral expressions. Thus, the collected interview responses are more closely linked to theoretically relevant personality traits and workplace behaviors. The collection of our dataset also combines validated self-report questionnaires with multi-perspective observer ratings. In general, our dataset simultaneously supports personality assessment and workplace-related competency analysis by providing psychometrically grounded self-reports and expert-rated behavioral annotations within a psychological-grounded assessment framework.

\section{Dataset Construction}
% Participants, Task Design, Annotation, Data Statistics, Reliability/Correlation

\subsection{Participants}

Participants were recruited via Prolific \cite{Peer2017}. Of the 793 recruited participants, we excluded those with incomplete responses ($n = 40$), no data-sharing consent ($n = 10$), failed attention checks ($n = 7$), abnormal HEXACO response patterns ($n = 6$; \cite{Barends_deVries_2019}), self-reported low effort ($n = 12$), corrupted audio ($n = 51$), or non-compliance identified by personality raters ($n = 21$). This resulted in a final sample of 646 participants.

Participants were native English-speaking adults living in the United States. The final sample was balanced by gender, including 309 men, 309 women, and 28 non-binary participants. Most participants identified as White ($n = 467$), followed by Black or African American ($n = 73$), Hispanic or Latino ($n = 46$), Asian ($n = 30$), other ethnicities ($n = 25$), or undisclosed ethnicity ($n = 5$). Participants had a mean age of 36.69 years ($SD = 11.92$) and an average of 15.96 years of work experience ($SD = 11.36$). Their education levels ranged from below high school to doctoral degrees. The study took approximately 60 minutes, and participants received \pounds10.00 as compensation.

\subsection{Interview question development \label{sec:questions}}
The interview followed a structured format, since previous literature suggests that structured (vs. non- or semi-structured) interviews have stronger reliability and validity\cite{Huffcutt2014}. Two types of questions were asked in the interview: 1) \textbf{generic questions} and 2) \textbf{personality questions}. Generic interview questions were designed to simulate commonly used selection interviews and elicit broad information about candidates that is relevant to overall interview evaluation and job-related competencies. In contrast, personality questions were developed based on \textit{Trait Activation Theory }(TAT) \cite{tett2003personality} and aimed to activate specific HEXACO personality traits through trait-relevant situations. This design enables the dataset to support both personality assessment and job-related competency assessment while improving the construct validity of the collected responses. 

The order of questions was fixed. Participants started with the generic questions and proceeded to the personality questions. Table \ref{tab:question} shows the six questions we developed as well as the order and corresponding personality traits. Below we describe the development and validation procedure for each type of interview question.

\subsubsection{\textbf{Generic questions}}An initial pool of 86 job interview questions taken from previous literature (e.g.,\cite{Hoque2013, Naim2015, Suen2019}) were selected. We screened those questions and included if they were (a) open-ended, (b) conveyed personality information to some extent, and (c) could apply to multiple jobs. Questions were excluded if they described 1) specific behaviors, 2) specific jobs , or 3) knowledge, values, and motives. This procedure ended up in retaining 61 questions.

To assess those 61 questions, we asked 17 professional recruiters from a Dutch consultancy company to assess how frequently they use each of those questions in practice. The recruiters had 10.38 years of experience on average and all of them held a Master’s degree (one held a PhD). Responses were given on a 3-point scale. The inter-rater agreement between the recruiters was $ICC(2,17) = 0.88$. After that, we asked four personality experts to assess each interview question on a 7-point scale using three criteria. The inter-rater agreement among the personality experts was $ICC(2,4) = 0.66$. Then, we calculated the average score per criterion (professional recruiters, personality experts) and excluded all questions that scored below the average (per criterion). At last, we selected two questions that received the highest ratings from recruiters and personality experts.

\begin{table*}[h!]
\caption{The content, order and type of the interview questions and their corresponding personality traits.}
\label{tab:question}
\begin{tabular}{@{}cccc@{}}
\toprule
Order & Interview question& Question Type & Personality Trait \\ \midrule
1     & \begin{tabular}[c]{@{}c@{}}What would you consider among your greatest \\ strengths and weaknesses as an employee?\end{tabular}                                                                                                          & Generic       & \textbackslash{}           \\ \midrule
2     & How would your best friend describe you?                                                                                                                                                                                                 & Generic       & \textbackslash{}           \\ \midrule
3     & \begin{tabular}[c]{@{}c@{}}Think of situations when you made professional decisions that \\ could affect your status or how much money you make. How do you usually \\ behave in such situations? Why do you think that is?\end{tabular} & Personality   & Honesty-Humility           \\ \midrule
4     & \begin{tabular}[c]{@{}c@{}}Think of situations when you joined a new team of people. \\ How do you usually behave when you enter a new team? \\ Why do you think that is?\end{tabular}                                                   & Personality   & Extraversion               \\ \midrule
5     & \begin{tabular}[c]{@{}c@{}}Think of situations when someone annoyed you. How do you usually \\ react in such situations? Why do you think that is?\end{tabular}                                                                          & Personality   & Agreeableness              \\ \midrule
6     & \begin{tabular}[c]{@{}c@{}}Think of situations when your work or workspace were not very organized. \\ How typical is that of you?  Why do you think that is?\end{tabular}                                                             & Personality   & Conscientiousness          \\ \bottomrule
\end{tabular}
\end{table*}

\subsubsection{\textbf{Personality questions}} For the development of personality interview questions, we created an initial pool of 25 past behavior interview questions for the personality traits of Honesty-Humility (\textit{n = 6}), Extraversion (\textit{n = 8}), Agreeableness (\textit{n = 5}), and Conscientiousness (\textit{n = 6}). The questions were developed to target the core facets of each personality trait. Questions were developed in a past-behavior format (e.g., “Think of situations when…”) since this type of format is more suited to elicit personality-relevant information according to previous research \cite{Levashina2014}.

Four personality experts independently selected one question per personality trait and later discussed any disagreements between them until a consensus was reached, retaining one question per personality trait. After some further editing, we ended up with four personality-related questions (Table \ref{tab:question}).

\subsection{Interview video collection}
Participants applied to a fictitious management traineeship position. The task is to complete an AVI (using a platform we developed for the purpose of the study), a personality questionnaire, a cognitive ability test, and to provide demographic information. During the AVI, participants responded to 2 generic and 4 personality questions designed in section \ref{sec:questions}. Participants were instructed to reply within 1-2 minutes to the interview questions. If their responses were too short (< 30 seconds), they were prompted to say something more, whereas if their responses were too long (> 150 seconds), they were prompted to conclude. Participants could record their responses only once, and there was no option to delete or resubmit a response.

\subsection{Self-report collection}
\subsubsection{\textbf{Self-reported personality traits}} Participants completed the HEXACO-100 inventory \cite{Lee2018}, which measures the six personality domains of Honesty-Humility, Emotionality, Extraversion, Agreeableness, Conscientiousness, and Openness to Experience. The questionnaire consists of 100 items rated on a 5-point Likert scale ranging from 1 (Strongly disagree) to 5 (Strongly agree). Domain scores were calculated by averaging the corresponding items. The HEXACO model was selected because it provides a comprehensive and well-validated framework for personality assessment and has demonstrated strong predictive validity for workplace behaviors and job performance. Cronbach’s alphas ranged from 0.84 (Honesty-
Humility and Openness to Experience) to 0.90 (Extraversion).

\subsubsection{\textbf{Cognitive ability}} To assess cognitive ability, participants completed the International Cognitive Ability Resource (ICAR) test \cite{Condon2014}. The ICAR consists of multiple cognitive tasks that measure general cognitive abilities, including verbal reasoning, letter and number series, matrix reasoning, and three-dimensional rotation. Participants' responses were aggregated to obtain an overall cognitive ability score. Cognitive ability was included because it is one of the most robust predictors of job performance and is frequently used in personnel selection research. Together, the HEXACO and ICAR measures provide validated self-report and cognitive assessments that support the analysis of personality traits, job-related competencies, and interview performance within AVI-Personality dataset. Failing to provide an answer within the time limit or leaving an item unanswered counted as a wrong answer. The general cognitive ability (g score) was the sum of all correct answers across the 28 items of the four scales. The average g score was 14.51 (range from 0 to 28) and its internal consistency was high (\textit{Cronbach’s alpha} = 0.82).

\subsection{Expert annotation}
\subsubsection{\textbf{Personality Traits}} Observer reports of personality were provided by a group of 12 raters. Raters followed a 9-hour training as described in \cite{Koutsoumpis2024}. Due to the high workload, each participant was rated by at least 3 personality psychologists. To test the Trait Activation Theory \cite{tett2003personality}, the rater annotated the personality traits based on two types of questions:

\begin{itemize}
\item \textbf{Six personality traits based on two \textit{generic questions}}:
raters watched responses to the two generic interview questions and evaluated all six HEXACO personality traits (i.e., Honesty-Humility (\textbf{H}), Emotionality (\textbf{E}), eXtraversion (\textbf{X}), Agreeableness (\textbf{A}), Conscientiousness (\textbf{C}), and Openness to experience (\textbf{O})). These questions were not specifically designed to activate any particular personality trait.

\item \textbf{Four personality traits based on four \textit{personality questions}}:
raters watched responses to the four personality questions and evaluated the corresponding personality traits (i.e., \textbf{H, X, A, C}). These questions were specifically designed to activate trait-relevant behaviors according to Trait Activation Theory \cite{tett2003personality}.
    
\end{itemize}

Previous multimodal personality datasets often relied on task-free or weakly structured content, making it difficult to determine whether observed behaviors genuinely reflected underlying personality traits. By collecting personality ratings from both generic interview questions and theory-driven personality questions, the dataset enables a comparison between non-activated and activated assessment conditions. The availability of both rating types also provides an opportunity to study how interview context influences personality expression and observer judgments in AVI settings.

Raters annotated the participants based on behaviorally anchored rating scales (BARS) \cite{zhang2025assessing}, which define rating levels through concrete behavioral descriptors and are commonly used to improve the structure and consistency of interview-based evaluations. The ratings were given on a 5-point scale (1 = Very low, 5 = Very high).

\subsubsection{\textbf{Job-related competencies}} The job-related competencies were provided by a group of five professional recruiters. The raters had on average 2.2 years of working experience. Due to the high workload, each participant received ratings from at least 2 raters. Raters assessed four job competencies and one overall interview performance score after having watched all six interview questions (raters manually rotated the order of questions to avoid ordering effects). The four job competencies (and their definitions) were taken from the manual of a Dutch consultancy company and were the following: 
\begin{itemize}
    \item \textit{\textbf{Integrity (Int)}}: The extent to which the candidate inspires trust, displays integrity in their interaction with others, treats others fairly, and adheres to high ethical standards;
    \item \textit{\textbf{Collegiality (Col)}}: The extent to which the candidate is open to and shows an interest in others and is willing to adapt one’s own activities to help others in their work;
    \item \textit{\textbf{Social versatility (Sv)}}: The extent to which the candidate has the ability to adapt one’s own behavior in a wide range of social situations in order to function effectively in different types of companies; 
    \item \textit{\textbf{Development orientation (Do)}}: The extent to which the candidate is willing to exert oneself in order to broaden and deepen knowledge and skills and to gain new experiences in order to grow professionally and increase the quality of one’s own work; 
    \item \textit{\textbf{Overall interview performance (Ip)}} : The extent to which the candidate would be able to fulfil the requirements of the management traineeship position.  
    
\end{itemize}

The ratings were provided using a BARS and given on a 5-point scale (1 = Very low; 5 = Very high), allowing to register up to one decimal point (e.g., ‘3.2’ was possible). We performed an exploratory principal component analysis (PCA) to test whether the four job competencies together with the overall interview performance score loaded on a single or multiple components. The results of the PCA showed that the five variables loaded on a single component. This finding suggests that the four job-related competencies and the overall interview performance score reflect a common underlying construct of interview effectiveness.

%The inter-rater agreement was $ICC_{1,2} = .55$ (for the sub-sample of n = 149 participants who received three interview performance ratings, $ICC_{2,3} = .60$).

\section{Dataset Analysis and Validation}
This section analyzes and validates the reliability, validity, fairness, and theoretical consistency of AVI-Personality dataset. For AVI-based assessment, high-quality annotations are as important as multimodal input data since unreliable or biased labels may limit both model development and practical applicability. Thus, we evaluate the inter-rater reliability of observer-reported personality traits and job-related competencies to test whether trained annotators provide consistent ratings. We then examine the validity of the annotations through self-other agreement for personality traits and internal nomological association for job-related competencies. We also analyze potential demographic bias and investigate the predictive relationship between personality traits and job-related competencies. This analysis examines whether the dataset captures theoretically expected links between individual dispositions and workplace-relevant behaviors.
\subsection{The reliability of observer-reports}
To test the reliability of observer-rated personality traits and job-related competencies, we conducted a mixed-effects reliability analysis using all observer ratings. For each personality trait and job-related competency, we fitted a linear mixed-effects model in which the rating score was predicted by a fixed intercept, with participant and rater specified as random effects. The participant-level variance captures true between-participant differences, whereas the rater-level variance captures systematic differences in rater severity and the residual variance captures unexplained rating error. Based on these variance components, we calculated two reliability indices: 
\begin{enumerate}
    \item \textbf{Mixed absolute agreement reliability} (\textit{mixed absolute ICC, \textbf{MAICC}}): treats systematic rater differences as measurement error;
    \item \textbf{Mixed consistency reliability} (\textit{mixed consistency ICC, \textbf{MCICC}}): focuses on whether raters rank participants in a similar way after rater-level differences are accounted for.
\end{enumerate}

\begin{table}[h!]
\centering
\caption{The mixed absolute ICC (MAICC) and (mixed consistency ICC (MCICC) of observer-reported personality traits and job-related competencies}
\label{tab:reliability}
\begin{tabular}{@{}cccc@{}}
\toprule
\multicolumn{2}{c}{}                                                                                                                                                                                                           & \textit{MAICC} & \textit{MCICC} \\ \midrule
\multicolumn{1}{c}{\multirow{6}{*}{\begin{tabular}[c]{@{}c@{}}\\ \\ Personality\\       traits from \\      generic \\      questions\end{tabular}}}     & Honesty-Humility                                                          & 0.533 & 0.715 \\ \cmidrule(l){2-4} 
\multicolumn{1}{c}{}                                                                                                                               & Emotionality                                                              & 0.633 & 0.779 \\ \cmidrule(l){2-4} 
\multicolumn{1}{c}{}                                                                                                                               & eXtraversion                                                              & 0.763 & 0.852 \\ \cmidrule(l){2-4} 
\multicolumn{1}{c}{}                                                                                                                               & Agreeableness                                                             & 0.639 & 0.782 \\ \cmidrule(l){2-4} 
\multicolumn{1}{c}{}                                                                                                                               & Conscientiousness                                                         & 0.706 & 0.821 \\ \cmidrule(l){2-4} 
\multicolumn{1}{c}{}                                                                                                                               & Openness                                                                  & 0.715 & 0.826 \\ \midrule
\multicolumn{1}{c}{\multirow{4}{*}{\begin{tabular}[c]{@{}c@{}}Personality\\       traits from \\      personality \\      questions\end{tabular}}} & Honesty-Humility                                                          & 0.654 & 0.792 \\ \cmidrule(l){2-4} 
\multicolumn{1}{c}{}                                                                                                                               & eXtraversion                                                              & 0.817 & 0.881 \\ \cmidrule(l){2-4} 
\multicolumn{1}{c}{}                                                                                                                               & Agreeableness                                                             & 0.693 & 0.814 \\ \cmidrule(l){2-4} 
\multicolumn{1}{c}{}                                                                                                                               & Conscientiousness                                                         & 0.747 & 0.844 \\ \midrule
\multicolumn{1}{c}{\multirow{5}{*}{\begin{tabular}[c]{@{}c@{}}\\ \\ \\ Job\\      related \\      competencies\end{tabular}}}                               & Integrity                                                                 & 0.324 & 0.525 \\ \cmidrule(l){2-4} 
\multicolumn{1}{c}{}                                                                                                                               & Collegiality                                                              & 0.368 & 0.566 \\ \cmidrule(l){2-4} 
\multicolumn{1}{c}{}                                                                                                                               & Social versatility                                                        & 0.359 & 0.558 \\ \cmidrule(l){2-4} 
\multicolumn{1}{c}{}                                                                                                                               & \begin{tabular}[c]{@{}c@{}}Development   \\      orientation\end{tabular} & 0.458 & 0.639 \\ \cmidrule(l){2-4} 
\multicolumn{1}{c}{}                                                                                                                               & \begin{tabular}[c]{@{}c@{}}Interview\\      performance\end{tabular}      & 0.379 & 0.575 \\ \bottomrule
\end{tabular}
\end{table}

As shown in Table \ref{tab:reliability}, observer-rated personality traits had moderate to high reliability. This indicates that even generic interview questions, which were not designed to activate a single specific trait, still provided sufficient behavioral information for trained raters to distinguish participants on personality traits. Among these traits, eXtraversion showed the highest reliability, suggesting that expressive and socially salient behaviors are especially observable in AVI responses. For ratings based on personality-targeted questions, reliability is higher than that of generic questions. The high reliability suggests that trait-activating questions successfully elicited more diagnostic behavioral cues for several personality dimensions. This pattern is consistent with the Trait Activation Theory \cite{tett2003personality}: situations that provide trait-relevant cues are more likely to reveal stable individual differences.

The reliability estimates for job-related competencies were lower than those for personality traits. These results suggest that recruiters showed only modest absolute agreement when evaluating competency levels, but achieved moderate consistency in ranking candidates. However, the higher MCICC indicate that recruiters shared a common basis for distinguishing candidates, especially for Development Orientation and overall interview performance.

Our findings support the reliability of the observer-rating procedure. The separation between generic and personality-targeted questions further shows that both types of AVI questions can produce reliable observer ratings. The high consistency reliability (i.e., MCICC) indicates that the ratings capture stable rank-order differences between participants rather than purely idiosyncratic rater impressions. Therefore, averaging ratings across multiple observers is appropriate for constructing criterion scores for AI-based analysis. Aggregation reduces the influence of individual rater severity and random rating error, while preserving the shared variance that reflects participants’ observable personality and competency-related behaviors. Thus, the averaged observer ratings provide a psychometrically defensible ground truth for training and evaluating AI models in automated personality and job-related competency assessment.

\subsection{Self-other agreement of personality traits}

\begin{figure}[h!]
    \centering
    \includegraphics[width=1\linewidth]{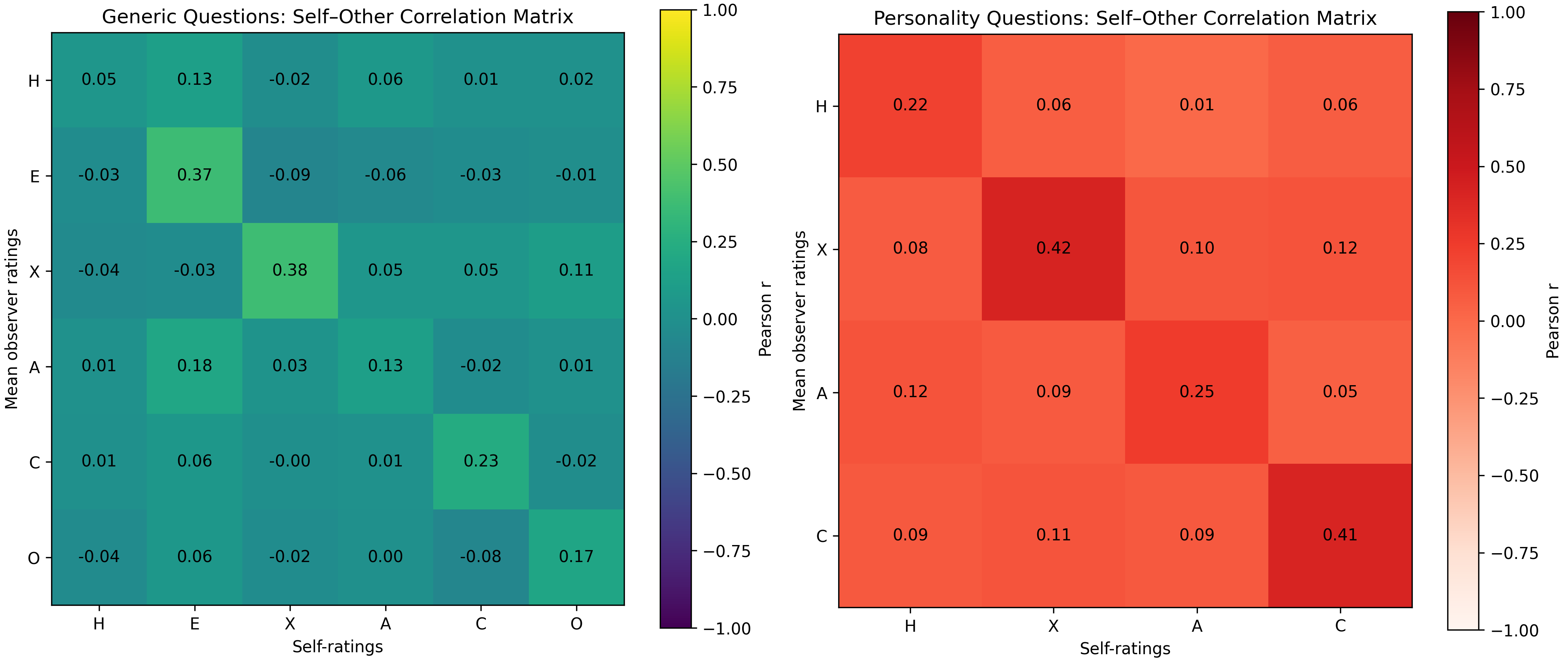}
    \caption{Self-other agreement of personality traits based on the observer reports from generic and personality questions}
    \label{fig:self-others}
\end{figure}

We further examined the construct validity of observer-rated personality traits by comparing the self-other agreement between participants' self-reported personality scores and mean observer ratings. Observer ratings were analyzed for generic questions and personality questions respectively. As shown in Figure \ref{fig:self-others}, the correlations based on generic questions were generally positive but varied across traits. The strongest self-other agreement was observed for eXtraversion ($r = 0.38$) and Emotionality ($r = 0.37$). This suggests that generic interview questions can capture some observable personality-relevant information, particularly for traits that are more behaviorally expressive or socially salient.

For personality-targeted questions, self-other agreement was stronger for most of the targeted traits. eXtraversion ($r = 0.42$) and Conscientiousness ($r = 0.41$) showed the highest agreement, followed by Agreeableness ($r =0.25$) and Honesty-Humility ($r = 0.22$). Compared with generic questions, personality questions produced higher self-other agreement, which indicates that trait-activating interview questions activated more trait-related behavioral cues that were aligned with participants' self-reported personality traits. These findings provide additional support for the construct validity of the observer ratings and are consistent with Trait Activation Theory \cite{tett2003personality}, which suggests that trait-relevant situations are more likely to reveal stable individual differences.

\subsection{Internal nomological association of Job-related Competencies}

To validate the nomological association of the job-related competency ratings, we conducted a multiple regression analysis using four job-related competencies (i.e., integrity, collegiality, social versatility and development orientation) to predict the overall interview performance (i.e., the extent to which a candidate was judged as suitable for the management traineeship position). The significant prediction by the competency ratings can indicate that these ratings capture practically relevant behavioral information for personnel selection.

\begin{table}[h!]
\centering
\caption{The multiple regression results for job-related competencies and interview performance}
\label{tab:mlr}
\begin{tabular}{@{}cccccc@{}}
\toprule
Predictor                                                          & $B$     & $SE$    & $\beta$  & $t$      & $p$     \\ \midrule
Int                                                          & 0.189 & 0.026 & 0.139 & 7.405  & <0.001 \\  \midrule
Col                                                       & 0.257 & 0.027 & 0.239 & 9.700  & <0.001 \\ \midrule
Sv                                                 & 0.461 & 0.029 & 0.405 & 16.119 & <0.001 \\  \midrule
Do & 0.352 & 0.024 & 0.296 & 14.829 & <0.001 \\ \bottomrule
\end{tabular}
\end{table}

As shown in Table \ref{tab:mlr}, the overall regression model was significant, ($F=864.33$, $p<0.001$, $R^2=0.871$ and adjusted $R^2=0.870$) explaining a large proportion of variance in interview performance. All four job-related competencies showed significant positive associations with interview performance. Among these competencies, social versatility was the strongest predictor ($\beta=0.405$). These results indicate that the four job-related competency ratings have high internal nomological association. Candidates who were rated high in these competencies were also more likely to receive higher interview performance evaluations. The large proportion of explained variance suggests that these competencies capture the core behavioral information underlying recruiters' overall hireability in AVIs.

\subsection{Predictive relationship between personality traits and job-related competencies}

To examine whether the personality annotations capture work-relevant behavioral information, we further investigated the predictive relationship between personality traits and job-related competencies. Specifically, we compared three sources of personality predictors using multiple regression analyses. The outcome variables for the analyses are five job-related competencies. For each outcome, we estimated three separate models using personality-question observer ratings, generic-question observer ratings, and self-reported personality scores as predictors. All continuous variables were standardized before analysis.

\begin{figure}[h!]
    \centering
    \includegraphics[width=1\linewidth]{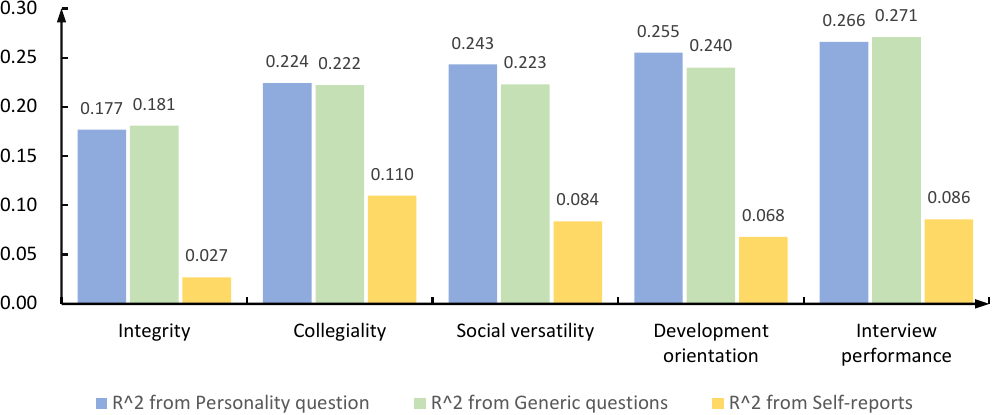}
    \caption{The $R^2$ of multiple regression analyses using the ratings from personality questions ($R_P^2$), generic questions ($R_G^2$) and self-reports ($R_S^2$)}
    \label{fig:mr_p_c}
\end{figure}

As shown in Figure \ref{fig:mr_p_c}, the observer-rated personality traits consistently demonstrated stronger predictive validity than self-reported personality traits. Models based on personality-question observer ratings explained 17.7\% to 26.6\% of the variance in job-related competencies and overall hireability, while models based on generic-question observer ratings explained 18.1\% to 27.1\% of the variance. In contrast, self-reported personality models explained only 2.7\% to 11.0\% of the variance. These results suggest that personality expressions observed in AVI responses are more strongly associated with workplace-related evaluations than self-reported personality scores.

In general, the results provide evidence for the construst validity of the observer-rated personality annotations in our dataset. Compared with self-reports, observer-rated personality traits derived from interview responses showed stronger associations with job-related competencies and overall interview performance. This supports the value of using structured AVI responses, particularly Trait Activation Theory-guided personality questions, to activate behavioral evidence relevant to both personality assessment and workplace competency evaluation.

\subsection{Fairness}
To examine demographic fairness, we tested whether observer-report personality and job-related competencies were systematically associated with demographic variables (i.e., Age, gender, education, and ethnicity). Separate regression models were fitted for each rating outcome. Specifically, age was treated as a continuous predictor, education as an ordered predictor, and gender and ethnicity as categorical predictors. Sparse ethnicity categories were combined into an underrepresented group. For each model, we tested the omnibus effect of each demographic variable and reported FDR-adjusted p-values and partial eta-squared effect sizes.

\begin{table}[h!]
\centering
\caption{Demographic fairness analysis for self-reported personality, observer-rated personality and job-related competencies. For each rating outcome, age, gender, education, and ethnicity were entered as demographic predictors in separate regression models. Values indicate partial eta-squared $(\eta_p^2)$ from omnibus demographic tests. Boldfaced values indicate significant demographic effects after $FDR$ correction ($^{*}p_{FDR} < .05$, $^{**}p_{FDR} < .01$, $^{***}p_{FDR}< .001$). Larger values indicate stronger demographic sensitivity of the corresponding rating outcome.}
\label{tab:fairness}
\tabcolsep=1mm
\begin{tabular}{@{}cccccc@{}}
\toprule
\textbf{}                                                                                                           & \textbf{} & \textbf{Gender}  & \textbf{Age}     & \textbf{Education} & \textbf{Ethnicity} \\ \midrule
\multirow{4}{*}{\textbf{\begin{tabular}[c]{@{}c@{}}Observer \\ ratings \\ from\\ personality \\ questions\end{tabular}}}   & H         & \textbf{0.029*}  & \textbf{0.012*}  & 0.008              & 0.003              \\ \cmidrule(l){2-6} 
                                                                                                                    & X         & 0.002            & 0.008            & 0.002              & 0.005              \\ \cmidrule(l){2-6} 
                                                                                                                    & A         & 0.003            & 0.008            & 0.001              & 0.006              \\ \cmidrule(l){2-6} 
                                                                                                                    & C         & \textbf{0.015*}  & \textbf{0.015**} & 0.008              & 0.013              \\ \midrule
\multirow{6}{*}{\textbf{\begin{tabular}[c]{@{}c@{}}Observer \\ ratings\\ from \\ generic \\ questions\end{tabular}}}       & H         & \textbf{0.035**} & 0.001            & 0.006              & 0.000              \\ \cmidrule(l){2-6} 
                                                                                                                    & E         & \textbf{0.177**} & \textbf{0.034**} & 0.001              & 0.008              \\ \cmidrule(l){2-6} 
                                                                                                                    & X         & 0.007            & 0.000            & \textbf{0.024**}   & 0.017              \\ \cmidrule(l){2-6} 
                                                                                                                    & A         & \textbf{0.024**} & 0.006            & \textbf{0.011*}    & 0.009              \\ \cmidrule(l){2-6} 
                                                                                                                    & C         & \textbf{0.015*}  & 0.000            & \textbf{0.047**}   & 0.004              \\ \cmidrule(l){2-6} 
                                                                                                                    & O         & \textbf{0.057**} & 0.007            & \textbf{0.017**}   & \textbf{0.019*}    \\ \midrule
\multirow{6}{*}{\textbf{\begin{tabular}[c]{@{}c@{}}Self\\ reported\\ personality\end{tabular}}}                     & H         & \textbf{0.013*}  & \textbf{0.094**} & \textbf{0.032**}   & 0.004              \\ \cmidrule(l){2-6} 
                                                                                                                    & E         & \textbf{0.132**} & \textbf{0.029**} & 0.001              & 0.003              \\ \cmidrule(l){2-6} 
                                                                                                                    & X         & 0.002            & \textbf{0.018**} & 0.005              & 0.010              \\ \cmidrule(l){2-6} 
                                                                                                                    & A         & 0.009            & \textbf{0.009*}  & 0.006              & 0.008              \\ \cmidrule(l){2-6} 
                                                                                                                    & C         & 0.001            & \textbf{0.024**} & 0.000              & 0.001              \\ \cmidrule(l){2-6} 
                                                                                                                    & O         & \textbf{0.018*}  & \textbf{0.018**} & 0.001              & 0.017              \\ \midrule
\multirow{5}{*}{\textbf{\begin{tabular}[c]{@{}c@{}}Observer\\ rated \\ job\\ related \\ competencies\end{tabular}}} & Int       & \textbf{0.017*}  & 0.008            & \textbf{0.015**}   & 0.002              \\ \cmidrule(l){2-6} 
                                                                                                                    & Col       & \textbf{0.028**} & 0.001            & \textbf{0.023**}   & 0.006              \\ \cmidrule(l){2-6} 
                                                                                                                    & Sv        & \textbf{0.016*}  & 0.004            & \textbf{0.039**}   & 0.007              \\ \cmidrule(l){2-6} 
                                                                                                                    & Do        & 0.007            & 0.001            & \textbf{0.056**}   & 0.009              \\ \cmidrule(l){2-6} 
                                                                                                                    & Ip        & \textbf{0.024**} & 0.002            & \textbf{0.043**}   & 0.005              \\ \bottomrule
\end{tabular}
\end{table}

The demographic fairness analysis showed that demographic associations varied across rating sources. As shown in Table \ref{tab:fairness}, self-reported HEXACO scores were most consistently associated with age, with all six dimensions showing significant age effects after FDR correction. Generic-question observer ratings showed more frequent demographic associations, particularly with gender and education. In contrast, observer ratings from personality questions showed fewer significant demographic effects, suggesting lower demographic sensitivity. For job-related competencies, education and gender were the most consistent demographic predictors, whereas ethnicity showed limited effects after correction. These results indicate that demographic variation should be considered when using AVI-based observer ratings as ground-truth labels for automatic personality and job-related competency assessment.

\section{Benchmark}
\subsection{Task Formulation and baseline selection}
To benchmark existing AI-based methods for assessing personality traits in AVIs, we conducted a comprehensive benchmark on our dataset. For the baseline benchmark, we use observer-rated personality traits from personality questions as the ground truth labels for training and testing. Compared with self-reported personality ratings, observer ratings better capture externally manifested personality expressions, which are the primary signals available to AI models from audiovisual interview data. We use observer ratings from personality questions instead of generic questions because trait-relevant situations facilitate the expression of personality variance. Thus, personality questions are supposed to elicit behavioral cues and provide more reliable labels for AI-based personality assessment.

Four categories of baselines (i.e., text-based, audio-based, visual-based, and multimodal-based methods.) are selected for dataset benchmark. Previous studies on AVI based personality assessment \cite{Koutsoumpis2024,zhang2024can,ghassemi2023unsupervised} have shown that textual responses often provide strong personality-relevant cues. Thus, it is worthwhile exploring whether incorporating additional behavioral modalities can further improve personality assessment beyond text-based approaches. 

For text-based methods, we included both traditional deep learning models and large language models (LLMs, i.e., DeepSeek-R1 \cite{guo2025deepseek} and GPT-4 \cite{OpenAI2023GPT4TR}) that have been applied to personality understanding from textual responses. Longformer \cite{hussain2025personality} T5-small \cite{lim2025persona} and Personality BERT \cite{jain2022personality} were selected as transformer-based architectures for capturing semantic information from interview transcripts. We also included recent personality-oriented LLM approaches, including MoE-Personality \cite{zhang2026mixture}, Jezoid \cite{cui2025less}, and PersonalityLLM \cite{zhang2026personalityllm}, which are specifically designed to improve personality assessment through large language model adaptation or reasoning mechanisms.

For audio-based methods, we included both handcrafted acoustic features and pretrained speech representation models. Traditional features, including eGeMAPS \cite{eyben2015geneva} and IS13-ComParE \cite{deb2017analysis}, were selected as widely used baselines in affective computing, while Whisper \cite{radford2023robust}, Wav2Vec2 \cite{baevski2020wav2vec}, Emotion2Vec \cite{ma2024emotion2vec}, and Kimi-Audio \cite{ding2025kimi} represent recent speech foundation models for capturing high-level vocal and affective cues.

For visual-based methods, we selected  models covering facial analysis, video representation learning, and vision-language understanding. DAN \cite{wei2017deep}, VAT \cite{girdhar2019video}, Swin Transformer \cite{liu2021swin}, ViT-MAE \cite{he2022masked}, SigLIP2 \cite{tschannen2025siglip}, Kimi-VL \cite{team2025kimi}, and Qwen3-VL \cite{bai2025qwen3} were included to evaluate the capability of different visual modeling paradigms in capturing personality-related behavioral cues.

For multimodal methods, we included approaches that integrate information from multiple behavioral modalities, which is particularly relevant to AVI scenarios. ResNet+BERT \cite{ResNet,BERT} represents an early multimodal fusion strategy combining visual and textual representations. HFUT-VisionXL \cite{li2025traits}, EMMR \cite{hu2026EMMR}, CAS-MAIS \cite{yang2025enhancing}, AU-Personality \cite{zhang2026llmbased}, and AVI2025 \cite{zhang2025assessing} were selected as  state-of-the-art methods specifically designed for personality assessment or AVI analysis. 

\subsection{Dataset split and evaluation metrics}
The dataset was split into training (70\%, \textit{n = 452}), validation (10\%, \textit{n = 64}), and testing (20\%, \textit{n = 130}) sets. The dataset is divided at the subject level, which makes sure that videos from a single subject are assigned exclusively to one of the training, validation, or testing sets. We consider the distribution of gender, age, and working experience of the subjects when splitting the dataset. Thus, we employ joint sampling to ensure that these three sets maintain similar distributions of these demographic and experiential variables. All the baseline methods were trained on the training and validation set expect for the commercial LLMs (i.e., gpt-4 and deepseek-r1). The results we report for these two methods are zero-shot performance.

To evaluate the performance of each baseline, we use Mean Squared Error (MSE) as the evaluation metrics. MSE is a widely adopted metric in regression tasks that calculates the average squared differences between predicted and actual values. Since this metric was used by several previous works \cite{hu2026EMMR,zhang2026mixture,zhang2026llmbased} and grand challenge \cite{zhang2025assessing} for AVI based personality assessment, it is well-suited for evaluating the performance of models targeting continuous traits.

\subsection{Results}

\begin{table}[h!]
\centering
\begin{threeparttable}
\caption{Baseline performance on our dataset measured by MSE}
\label{tab:baseline}
\centering
\tabcolsep = 1mm
\begin{tabular}{@{}cccccc@{}}
\toprule
\multicolumn{6}{c}{Text-based methods}                          \\ \midrule
Traits           & H     & X     & A     & C     & \textbf{AVG} \\ \midrule
Longformer \cite{hussain2025personality}       & 0.270 & 0.470 & 0.090 & 0.170 & 0.250        \\
T5-small  \cite{lim2025persona}        & 0.123 & 0.268 & 0.192 & 0.213 & 0.199        \\
deepseek-r1  \cite{guo2025deepseek}     & 1.010 & 0.605 & 0.699 & 0.761 & 0.769        \\
gpt-4 \cite{OpenAI2023GPT4TR} & 0.441 & 0.735 & 0.555 & 0.821 & 0.638        \\
MoE-Personality \cite{zhang2026mixture}& 0.232 & 0.363 & 0.440 & 0.401 & 0.359        \\
Jezoid \cite{cui2025less} & 0.158 & \textbf{0.149} & 0.139 & \textbf{0.104} & 0.137        \\
Personality BERT  \cite{jain2022personality}  & 0.117 & 0.292 & 0.090 & 0.250 & 0.187        \\
PersonalityLLM \cite{zhang2026personalityllm}& \textbf{\underline{0.080}} & 0.251 & \textbf{\underline{0.059}} & 0.110 &\textbf{ 0.125}        \\ \midrule
\multicolumn{6}{c}{Audio-based methods}                         \\ \midrule
Whisper \cite{radford2023robust} & 0.186 & \textbf{0.217} & \textbf{0.204} & \textbf{0.154} & \textbf{0.190 }      \\
Emotion2Vec \cite{ma2024emotion2vec}     & 0.189 & 0.243 & 0.210 & 0.182 & 0.206        \\
Wav2Vec2 \cite{baevski2020wav2vec}        & 0.190 & 0.269 & 0.219 & 0.188 & 0.217        \\
eGeMAPS \cite{eyben2015geneva}   & 5.526 & 4.255 & 5.842 & 5.631 & 5.314        \\
is13-compare \cite{deb2017analysis}    & 3.045 & 4.309 & 2.871 & 2.967 & 3.298        \\
Kimi-audio \cite{ding2025kimi} & \textbf{0.162} & 0.450 & 0.343 & 0.257 & 0.303        \\ \midrule
\multicolumn{6}{c}{Visual-based methods}                        \\ \midrule
ViT-MAE \cite{he2022masked}         & 0.185 & 0.268 & 0.219 & 0.185 & 0.214        \\
SigLIP2 \cite{tschannen2025siglip}  & 0.185 & 0.268 & 0.219 & 0.185 & 0.214        \\
DAN \cite{wei2017deep} & 0.384 & 0.199 & 0.243 & 0.181 & 0.252        \\
VAT \cite{girdhar2019video}             & 0.223 & 0.309 & 0.332 & 0.375 & 0.310        \\
Swin-transformer \cite{liu2021swin} & \textbf{0.164} & \textbf{0.167} & \textbf{0.182} & \textbf{0.132} & \textbf{0.161}        \\
Kimi-VL \cite{team2025kimi}          & 0.353 & 0.443 & 0.244 & 0.287 & 0.332        \\
Qwen3-VL \cite{bai2025qwen3}        & 1.225 & 0.535 & 0.926 & 0.957 & 0.911        \\ \midrule
\multicolumn{6}{c}{Multimodal-based methods}                    \\ \midrule
ResNet\cite{ResNet}+BERT \cite{BERT}      & 0.256 & 0.297 & 0.528 & 0.315 & 0.349        \\
HFUT-VisionXL \cite{li2025traits}    & \textbf{0.126} & 0.130 & \textbf{0.136} & 0.099 & \textbf{\underline{0.123}}        \\
EMMR \cite{hu2026EMMR}             & 0.161 & 0.153 & 0.150 & \textbf{\underline{0.093}} & 0.139        \\
CAS-MAIS \cite{yang2025enhancing}        & 0.153 & 0.149 & 0.151 & 0.121 & 0.144        \\
AU-personality \cite{zhang2026llmbased}  & 0.156 & 0.138 & 0.160 & 0.108 & 0.140        \\
AVI2025 \cite{zhang2025assessing}         & 0.192 & \textbf{\underline{0.111}} & 0.229 & 0.188 & 0.180        \\ \bottomrule
\end{tabular}
\begin{tablenotes}
\small
\item[1] The MSE we reported for gpt-4 and deepseek-r1 are zero-short.
\item[2] AVG denotes the average MSE across traits. 
\item[3] \textbf{Bold values} mark the best result within each modality group.
\item [4] \underline{Underlined values} mark the best result across all baselines.
\item[5] For example, \textbf{\underline{0.093}} (EMMR) means the best Conscientious (C) results among multimodal-based methods methods (bold) and the best performance across all methods (underlined).
\end{tablenotes}
\end{threeparttable}
\end{table}

Table \ref{tab:baseline} reports the MSE of different baseline methods on our dataset. Among text-based methods, PersonalityLLM \cite{zhang2026personalityllm} obtained the best average results (also for H and A) for across all baselines. This indicates that trait-relevant verbal content provides strong cues for personality assessment, especially when LLMs are adapted to personality assessment tasks. In contrast, general-purpose LLMs such as gpt-4 \cite{OpenAI2023GPT4TR} and deepseek-r1 \cite{guo2025deepseek} achieved much higher MSE, which suggests that model scale alone is insufficient without task-specific alignment.

Audio and visual based methods provided useful but more limited unimodal information. Self-supervised speech models such as Whisper, Emotion2Vec \cite{ma2024emotion2vec}, and Wav2Vec2 \cite{baevski2020wav2vec}, outperformed handcrafted acoustic feature sets such as eGeMAPS \cite{eyben2015geneva} and IS13-ComParE \cite{deb2017analysis}. This indicates that learned speech representations are more effective than traditional acoustic descriptors for capturing personality-related vocal cues. For the visual-based methods, Swin Transformer \cite{liu2021swin} achieved the best performance across all four traits and the lowest average MSE among visual baselines. This suggests that modeling spatial and temporal visual patterns is important for extracting personality-related behaviors from video. However, compared with text-based methods, audio or visual only models generally produced higher MSEs, which indicates that nonverbal channels alone provide relatively limited information for personality assessment. The personality questions elicit rich trait-relevant semantic content, whereas vocal and visual behaviors are more indirect and noisy signals. Thus, they can reflect transient presentation style, affective state, or recording conditions rather than stable personality traits.

Multimodal methods achieved the best overall performance. However, the advantage of multimodal methods over strong text-based baselines was relatively small. The best results for individual traits were distributed among both text-based and multimodal-based methods. For example, EMMR \cite{hu2026EMMR} (multimodal-based) achieved the best result for C, whereas PersonalityLLM \cite{zhang2026personalityllm} (text-based) achieved the best result for both H and A. These findings suggest that multimodal fusion can improve overall robustness, while its effectiveness depends on how well complementary linguistic, acoustic, and visual cues are integrated. Thus, our benchmark highlights both the importance of textual information in trait-activating AVI responses and the need for more effective multimodal fusion strategies for personality assessment.

\section{Limitation and future works}
Although AVI-Personality provides a trait-activated multimodal dataset for personality and job-related competency assessment, several limitations still exist. First of all, the personality questions were designed for four job-related HEXACO traits while Emotionality and Openness were assessed through self-reports and observer ratings from generic questions. Future studies should design and validate personality questions for all HEXACO domains to allow a more complete examination of Trait Activation Theory in AVI-based assessment. 

In addition, although the observer ratings were provided by trained raters using BARS, the ratings can still contain rater-specific preferences or demographic sensitivity. The reliability results showed moderate to high consistency for observer-rated personality traits while lower absolute agreement for job-related competencies. This suggests that competency judgments may require additional calibration. Future work should examine the decision processes of human raters and investigate whether AI models trained on these labels reproduce or amplify demographic patterns present in human ratings.

At last, the benchmark provides a broad comparison across different AI model. However, it does not exhaust all possible modeling strategies. The results showed that text-based methods were highly competitive and  multimodal methods achieved the best average performance. However, the advantage of multimodal fusion over strong text-based baselines was relatively small. This indicates that simply combining modalities may be insufficient for AVI-based personality assessment. Future work should explore more effective multimodal fusion strategies to better fuse linguistic, acoustic, and visual behavioral cues for personality assessment.

\section{Conclusion}
Our work introduced AVI-Personality, a trait-activated multimodal dataset for personality and job-related competency assessment from asynchronous video interviews. The dataset contains 3,876 videos from 646 participants, with structured responses to generic and personality interview questions. It provides multimodal behavioral data together with self-reported personality traits, observer-rated personality traits, recruiter-rated job-related competencies, interview performance, and cognitive ability.

We validated the dataset through reliability, construct validity, internal nomological association, and fairness analyses. The results showed that observer-rated personality traits, especially those based on personality-targeted questions, had reliable and meaningful annotations. Job-related competencies significantly predicted interview performance. The benchmark further showed that text-based methods provide strong personality-relevant signals, while multimodal methods achieved the best overall performance. Overall, our dataset provides a psychometrically grounded resource for developing reliable, valid, and fair AI-based assessment algorithms.
\bibliographystyle{unsrt}  

\bibliography{AVI6refrence}

\end{document}